\documentclass[prb,reprint,aps,superscriptaddress,showpacs]{revtex4-1}

\usepackage{graphicx}
\usepackage{amssymb}
\usepackage{amsmath}
\usepackage{graphicx}
\usepackage{color}

\newcommand{\lmpo}{LiMnPO$_\mathsf{4}$}
\newcommand{\li}{$^7$Li}
\newcommand{\phos}{$^{31}$P}
\newcommand{\p}{$^{31}$P}

\begin{document}


\title{Experimental evidence for the coupling of Li-motion and structural distortions in LiMnPO$_4$}

\author{Christian Rudisch}
\author{Hans-Joachim Grafe}
\author{Jochen Geck}
\author{Sven Partzsch}
\affiliation{IFW Dresden, Leibniz Institute for Solid State and Materials Research, D-01171 Dresden,
Germany}
\author{M.~v.~Zimmermann}
\affiliation{Deutsches Elektronen-Synchrotron DESY, D-22603 Hamburg, Germany}
\author{Nadja Wizent}
\affiliation{IFW Dresden, Leibniz Institute for Solid State and Materials Research, D-01171 Dresden,
Germany}
\affiliation{Kirchhoff-Institut f\"ur Physik, Universit\"at Heidelberg, D-69120 Heidelberg, Germany}
\author{R\"udiger Klingeler}
\affiliation{Kirchhoff-Institut f\"ur Physik, Universit\"at Heidelberg, D-69120 Heidelberg, Germany}
\author{Bernd B\"uchner}
\affiliation{IFW Dresden, Leibniz Institute for Solid State and Materials Research, D-01171 Dresden,
Germany}
%Hamburger Synchrotronstrahlungslabor HASYLAB at Deutsches Elektronen-Synchrotron DESY,
%Notkestra\ss e 85, D-22603 Hamburg, Germany\\

%75.25.+z - Spin arrangements
%76.60.-k - Nuclear magnetic resonance and relaxation
%75.50.Ee - Antiferromagnetics
%61.10.Nz - X-ray diffraction
\begin{abstract}
We present a detailed $^7$Li- and $^{31}$P-NMR study on single
crystalline LiMnPO$_4$ in the paramagnetic and antiferromagnetic
phase~(AFM, $T_N \sim 34$~K).
This allows us to determine the spin directions in the field-induced spin-flop phase.
In addition, the anisotropic dipolar hyperfine coupling tensor of
the $^7$Li- and $^{31}$P-nuclei is also fully determined by
orientation and temperature dependent NMR experiments and compared
to the calculated values from crystal structure data. Deviations
of the experimental values from the theoretical ones are discussed
in terms of Mn disorder which is induced by Li-disorder. In fact,
the disorder in the Mn-sublattice is directly revealed by our
diffuse x-ray scattering data. The present results provide
experimental evidence for the Li-diffusion strongly coupling to
structural distortions within the MnPO$_4$ host, which is expected
to significantly affect the Li-mobility as well as the performance
of batteries based on this material.
\end{abstract}

% \begin{keyword}
% %% keywords here, in the form: keyword \sep keyword
% Nuclear Magnetic Resonance \sep
% Phosphates \sep
% Lithium compounds \sep
% Antiferromagnetism
% %% MSC codes here, in the form: \MSC code \sep code
% %% or \MSC[2008] code \sep code (2000 is the default)
%
% \end{keyword}

\date{\today}
\maketitle

\section{Introduction}

Since the pioneering work of Padhi et al.~\cite{Padhi} in 1997,
the phospho-olivine materials are highly interesting for an
application in rechargeable lithium batteries. LiMnPO$_4$ is a
member of the olivine type lithium phosphate family and features
several advantages regarding battery technology such as excellent
chemical and thermal stability, non-toxicity in contrast to
LiCoO$_2$, and economic viability and availability of the raw
materials~\cite{jahnteller2010,ecs2009,Guohua}.
Despite these clear advantages of LiMnPO$_4$ for battery
applications, there are also reports about its poor
electrochemical
performance\cite{Padhi,AsariPRB2011,Yamada2001A,Yamada2001B}. The
reasons for this currently remain unclear, but it has been found
that LiMnPO$_4$ incorporated in carbon coated nanostructures
provides a competitive next generation cathode material with a
stable reversible capacity up to 145 mAh/g and a rather flat
discharge voltage curve at 4.1 V \cite{Wang2009,Aravindan2013}.

Besides its relevance for battery applications \lmpo\/ also
exhibits interesting magnetic properties. More specifically,
applying a magnetic field leads to a spin flop phase with a slight
ferromagnetic canting, which could exhibit magnetoelectric effects
\cite{Toft-PetersenPRB2012}. In addition, in LiMPO$_4$ with M = Co
or Ni unusual ferrotoroidic domains have been discovered
\cite{Aken2007}.

In this article we address the origin of the poor electrochemical
performance of \lmpo\ as well as the possible magnetoelectric
effect in this material. To this end, we performed full
orientation and temperature dependent \li\ and \p\ NMR experiments
on a single crystal of \lmpo . This allowed us to determine the
complete dipolar hyperfine coupling tensor $\hat{A}_\mathrm{dip}$
which is compared to the calculated one from the positions of the
atoms in the unit cell. Deviations of the experimentally obtained
hyperfine coupling tensors from the calculated tensors and an
anomalous broadening of the resonance lines in the paramagnetic
phase are discussed in terms of Mn disorder in the samples. The
conclusions from NMR are corroborated by our diffuse x-ray
diffraction (HE-XRD) experiments, which directly reveal the
disorder in the Mn-sublattice. The NMR spectra in the AFM phase
have been measured at 4.2 K, and the number of resonance lines
agrees with the calculated spectra. Furthermore, the spectra
allowed us to determine the spin directions in the spin-flop phase
as well as the tilt of the spins due to the applied external
magnetic field, which is important in order to assess the possible
magnetoelectric effect in this material.

Our findings of Mn disorder in \lmpo\ provides firm experimental
evidence that the movement of Li within LiMnPO$_4$ is strongly
coupled to the lattice. Certainly, this affects also the mobility
of the Li in \lmpo , and therefore the performance of this
material as a battery. Interestingly, our results are perfectly
consistent with a recent theoretical study which found a formation
of a vacancy-polaron complex by a lithium vacancy and a
corresponding hole-polaron at the fully lithiated limit owing to
lattice distortion and Coulomb interaction between them
\cite{AsariPRB2011}, and thereby explaining the poor
electrochemical performance of \lmpo\ if non-carbon coated
microstructures or single crystals are considered.

\section{Experimental details}

\subsection{Sample preparation and characterization}

\begin{figure}[!t]
  \centering
  \includegraphics[width=0.40\textwidth]{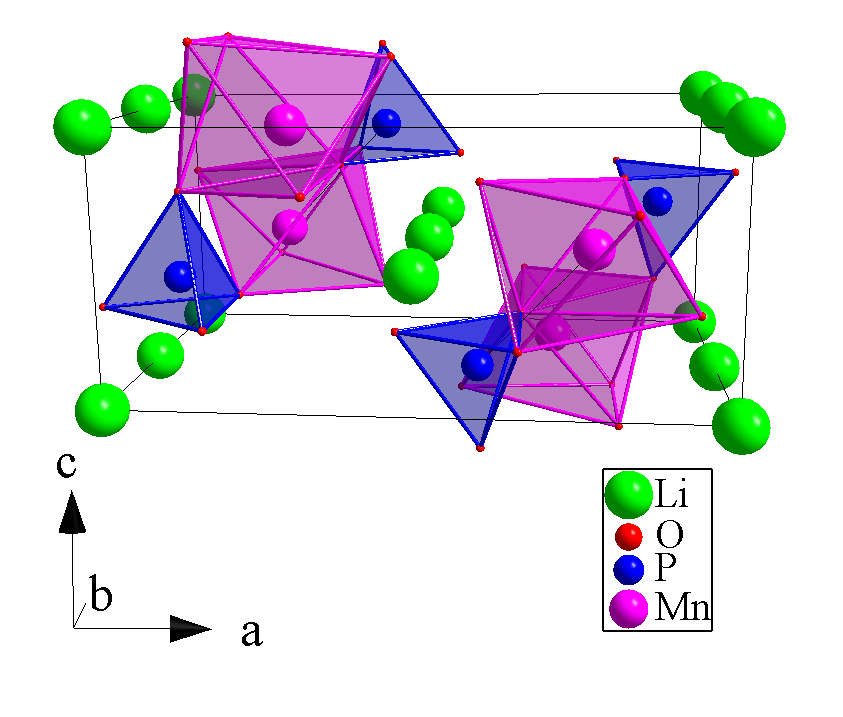}
  \caption{Crystal structure of LiMnPO$_4$. The crystallographic axes
  ($a$, $b$, and $c$) are along the edges of the unit cell. The slightly distorted
  MnO$_6$ octahedra (shaded in violet) share the oxygen anions with
  the PO$_4$ tetrahedra (shaded in blue).}
  \label{fig:crystal_structure}
\end{figure}

The single crystal investigated in this work was grown at the IFW
Dresden by the floating-zone method~\cite{wizent2009,wizent2011}.
A phase pure single crystalline grain of the approximate size of 2
$\times$ 2 $\times$ 4 mm was cut out of the grown rod. The phase
purity was checked by x-ray diffraction measurement. The crystal
structure of LiMnPO$_4$ shown in Fig.~\ref{fig:crystal_structure}
belongs to the Pnma spacegroup. The Mn atoms are surrounded by
distorted octahedra of oxygen atoms. The P atoms are tetrahedrally
surrounded by oxygen atoms forming PO$_{4}^{3-}$
groups~\cite{Mays1963}. There are four Li and four P positions
which are crystallographically equivalent. The Mn ions are in a
Mn$^{2+}$ state resulting in a spin of $S=5/2$ and an effective
magnetic moment of 5.9 $\mu_B$~\cite{wizent2009}. The Mn magnetic
moments align antiferromagnetically below $T_N \sim 34$~K
\cite{wizent2009} along the \textit{a}-axis \cite{LiPRB2009}.

\subsection{NMR and Hyperfine Coupling Tensor}

Nuclear Magnetic Resonance (NMR) is an ideal tool to study both
the Li diffusion process, which is important for application in
lithium ion batteries, as well as the local magnetic properties.
The Li diffusion process determines the mobility of the Li ions in
a material and can be investigated by \li-NMR linewidth and spin
lattice relaxation measurements~\cite{heitjans,nakamura2006},
while the magnetism can be probed by any nucleus in the material
that is coupled to the magnetic ion. For both processes, the
detailed knowledge of the hyperfine coupling tensor between the
electron spin and the nuclear probes --here the nuclear spins of
$^7$Li and $^{31}$P-- is helpful.

All NMR experiments reported here are done at a constant field of
$H_0=7.0494$~T. For orientation dependent NMR measurements a probe
with a single axis goniometer was used which allows to rotate the
sample by a certain angle inside the magnet. The temperature
dependent NMR measurements were done in a temperature range from
4.2~K up to 420~K. In the following, only the dipolar hyperfine
coupling will be described in detail, since as we will see the
other hyperfine couplings are either negligible (quadrupolar
coupling, diamagnetic and orbital shift) or isotropic (Fermi
contact from unpaired $s$-electrons on the \phos ).

The Hamiltonian describing the dipolar coupling of the nuclear
spin to the Mn electronic spins is expressed by
\begin{equation}
  \mathcal{H}_\mathrm{dip} = \hslash \gamma_n B_\mathrm{dip} = -\frac{\mu_0}{4 \pi} \gamma_e \gamma_n \hslash^{2}
\sum_j \vec{I} \cdot \hat{A}_{dip,j} \cdot \vec{S}^T_j.
  \label{eqn:interaction}
\end{equation}
Where $B_\mathrm{dip}$ represents the dipolar field of the
electronic spin located at the Mn site. $\gamma_n$ corresponds to
the gyromagnetic ratio of the nuclei $^7$Li and
$^{31}$P~($\gamma_{Li}$=16.5461~MHz, $\gamma_P$=17.2347~MHz).
$\gamma_e$ is the gyromagnetic ratio of the electron. The nuclear
and $j$th electron spin are represented by $\vec{I}$ and
$\vec{S}^T_j$ respectively. The anisotropic dipolar coupling
between the nuclear and the electron spins is described by the
dipolar hyperfine coupling tensor $\hat{A}_\mathrm{dip}$ (see e.g.
\cite{Abragam}). It has been calculated for the paramagnetic
$\hat{A}^{para}_\mathrm{dip}$ and AFM phase
$\hat{A}^\mathrm{AFM}_\mathrm{dip}$ from the crystal structure
data.

All electron spins within a sphere with a radius of 100~\AA{} have
been considered, leading to the same tensor elements $A_{mn}$ of
$\hat{A}_\mathrm{dip}$ as reported by Mays et al.~\cite{Mays1963}.
The different positions in the crystal of the $^7$Li and the
$^{31}$P naturally lead to different dipolar hyperfine coupling
tensors $\hat{A}_\mathrm{dip,Li}$ and $\hat{A}_\mathrm{dip,P}$,
respectively. Crystallographically, there are four equivalent
$^7$Li and $^{31}$P sites in the unit cell. However, for each
nucleus, the diagonal elements $A_{mm}$ of $\hat{A}_\mathrm{dip}$
are the same, whereas the off diagonal elements $A_{mn} (m\neq n)$
are different. Therefore, for both nuclei, single resonance lines
are expected for $H_0 \parallel a,b,$ and $c$, whereas the
different off diagonal elements lead to a splitting of the
resonance lines for angles off these high symmetry directions in
the paramagnetic phase.

In the antiferromagnetically ordered state,
$\hat{A}^\mathrm{AFM}_\mathrm{dip,Li}$ vanishes, i.e. no internal, dipolar field
remains at the Li sites, leading to an unshifted single resonance
line for the \li . The calculation of $\hat{A}^\mathrm{AFM}_\mathrm{dip,P}$
leads to four magnetically inequivalent P sites in the AFM phase.
Note that the diagonal elements of these P sites are different.
Therefore, up to four $^{31}$P-NMR lines are expected in the AFM
phase. The calculated tensor elements of the paramagnetic and the
AFM phase for $^7$Li and $^{31}$P nuclei are listed in
Tab.~\ref{tab:Tensors} where a comparison of calculated and
experimentally determined values is given.

\subsection{X-ray diffraction experiments}

Complementary structural information has been obtained by
high-energy x-ray diffraction (HE-XRD) studies on the same single
crystal. The experiments were performed at the beamline BW5 at the
HASYLAB in Hamburg, using x-rays with photon energies of 100~keV.
Due to the large penetration depths at this energy ($\sim$1~mm),
surface effects have no influence on the detected signal, ensuring
the detection of real bulk properties. We performed triple-axis
diffraction in horizontal Laue geometry utilizing the (111)
reflection of a Si/Ge-monochromator and Si/Ge-analyzer crystal.

\section{Results and Discussion}

\subsection{NMR spectra and temperature dependence}

\begin{figure}[!t]
  \centering
  \includegraphics[width=0.45\textwidth]{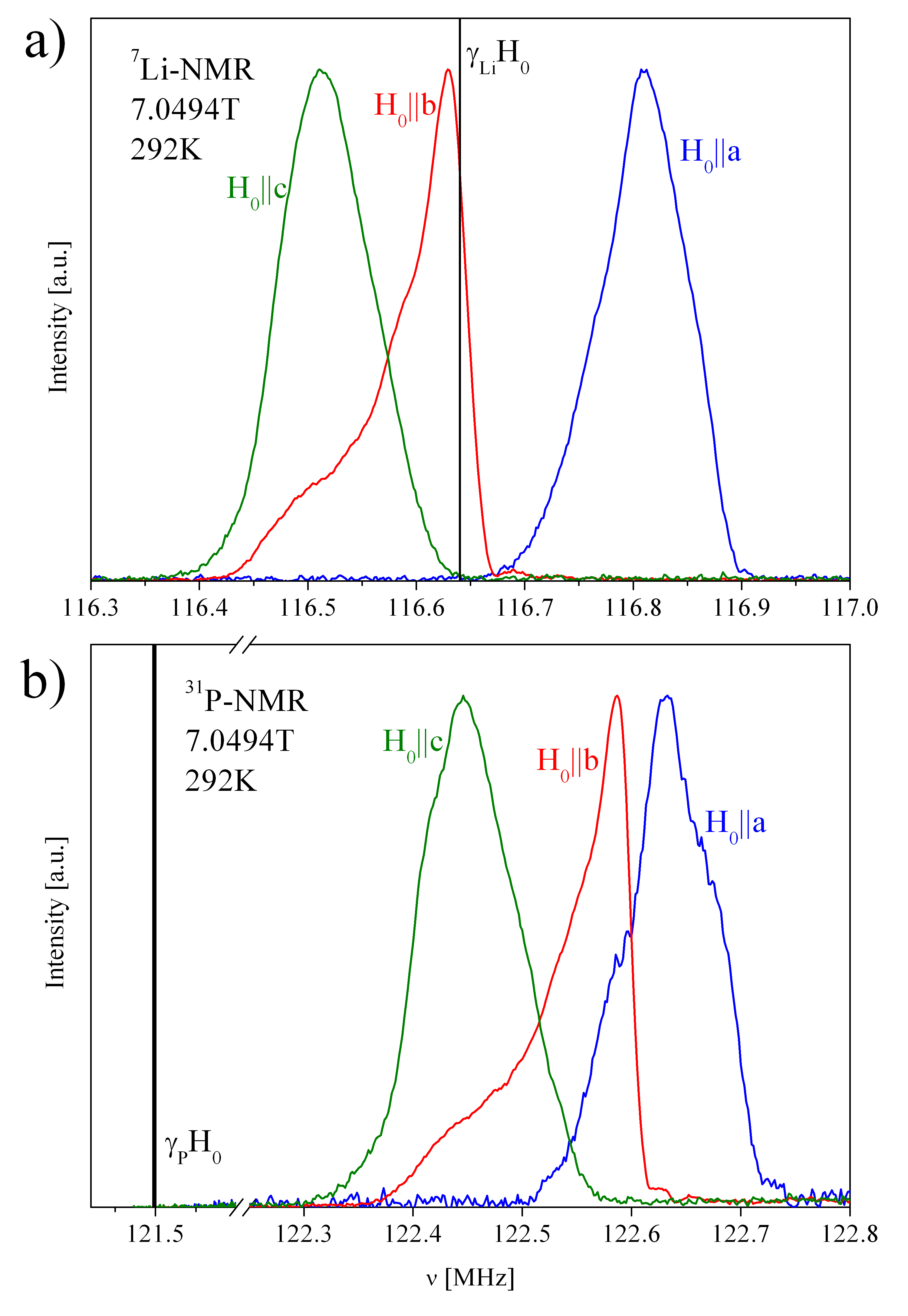}
  \caption{(a) $^7$Li-NMR spectra and (b) $^{31}$P-NMR spectra for the orientations $H_0\Arrowvert a$,
$H_0\Arrowvert b$ and $H_0\Arrowvert c$. The average linewidth is
$\sim$80~kHz. The $^7$Li-NMR spectra in (a) exhibit a small shift
around the resonance line of a bare nucleus~($\gamma_{Li}H_0$).
The $^{31}$P-NMR spectra in (b) are showing large shifts to higher
resonance frequencies which can be due to the isotropic contact
interaction, see text.}
  \label{fig:NMR_abc_Li_P}
\end{figure}

In Fig.~\ref{fig:NMR_abc_Li_P} the $^7$Li- and $^{31}$P-NMR
spectra are shown for $H_0\Arrowvert a$, $H_0\Arrowvert b$, and
$H_0\Arrowvert c$ at 292~K. Due to the anisotropy of the dipolar
hyperfine couplings $\hat{A}_\mathrm{dip,Li}$ and
$\hat{A}_\mathrm{dip,P}$ the resonance frequency $\nu$ is
different for the different orientations.
The $^7$Li-NMR spectra for all different orientations are spanning
a frequency interval of 500~kHz at 292~K. This value agrees well
with the full width of $^7$Li-NMR spectra in LiMnPO$_4$ powder
samples reported in \cite{comparative}. The linewidth shows a
linear field dependence~(e.g. $\Delta \nu$(7T)$~\sim$~80~kHz,
whereas $\Delta \nu$(3T)$~\sim$~40~kHz). This indicates a dipolar
broadening, where each $^7$Li ($^{31}$P) nucleus sees a slightly
different dipolar hyperfine field of the Mn electronic spins.
However, the asymmetric shape of the resonance lines~(especially
for $H_0\Arrowvert b$) differs much from the expected Gaussian
shape of a resonance line in a single crystal, indicating a
substantial amount of disorder in the sample. For this reason, we
have calculated the Knight shift of the \li\ and \p\ from the
center of gravity of the resonance lines, and the width $\Delta
\nu$ of the spectra has been determined by the square root of the
second moment.

Furthermore, a comparison of the resonance lines in
Fig.~\ref{fig:NMR_abc_Li_P} shows that for the corresponding
orientations the shape of $^7$Li- and $^{31}$P-NMR spectra are
very similar. This indicates that the shape of the resonance lines
of both nuclei must be affected by the same factor.
If only the \li\ is disordered in the crystal, this would not affect the
$^{31}$P resonance lines and vice versa.

A misalignment of the external field can be excluded, too, since
the crystal was accurately oriented by making use of the angle
dependence of the resonance frequencies of the $^7$Li- and
$^{31}$P resonance lines (see Fig.~\ref{fig:Li_P_ac},
\ref{fig:Li_P_ab}, and \ref{fig:Li_P_bc}). A quadrupolar
broadening or splitting of the $^7$Li- spectra ($I=3/2$) could
also not be observed, indicating a very small electric field
gradient at the Li site. Therefore, the anomalous shape of the
resonance lines of both nuclei must originate either from a
distribution of local moments of the Mn, i.e. a distribution of
the valence of the Mn, or a distribution of the hyperfine coupling
which can originate from a site disorder of the Mn. Since there
are no indications for a distribution of the Mn valence
\cite{wizent2009}, we conclude that there is substantial Mn site
disorder. A detailed comparison of the calculated and experimental
hyperfine coupling tensor elements in the following points into
the same direction. Further investigations by diffusive x-ray
scattering (see Chapter 4) corroborate the existence of Mn site
disorder in LiMnPO$_4$.

\begin{figure}[!t]
  \centering
  \includegraphics[width=0.45\textwidth]{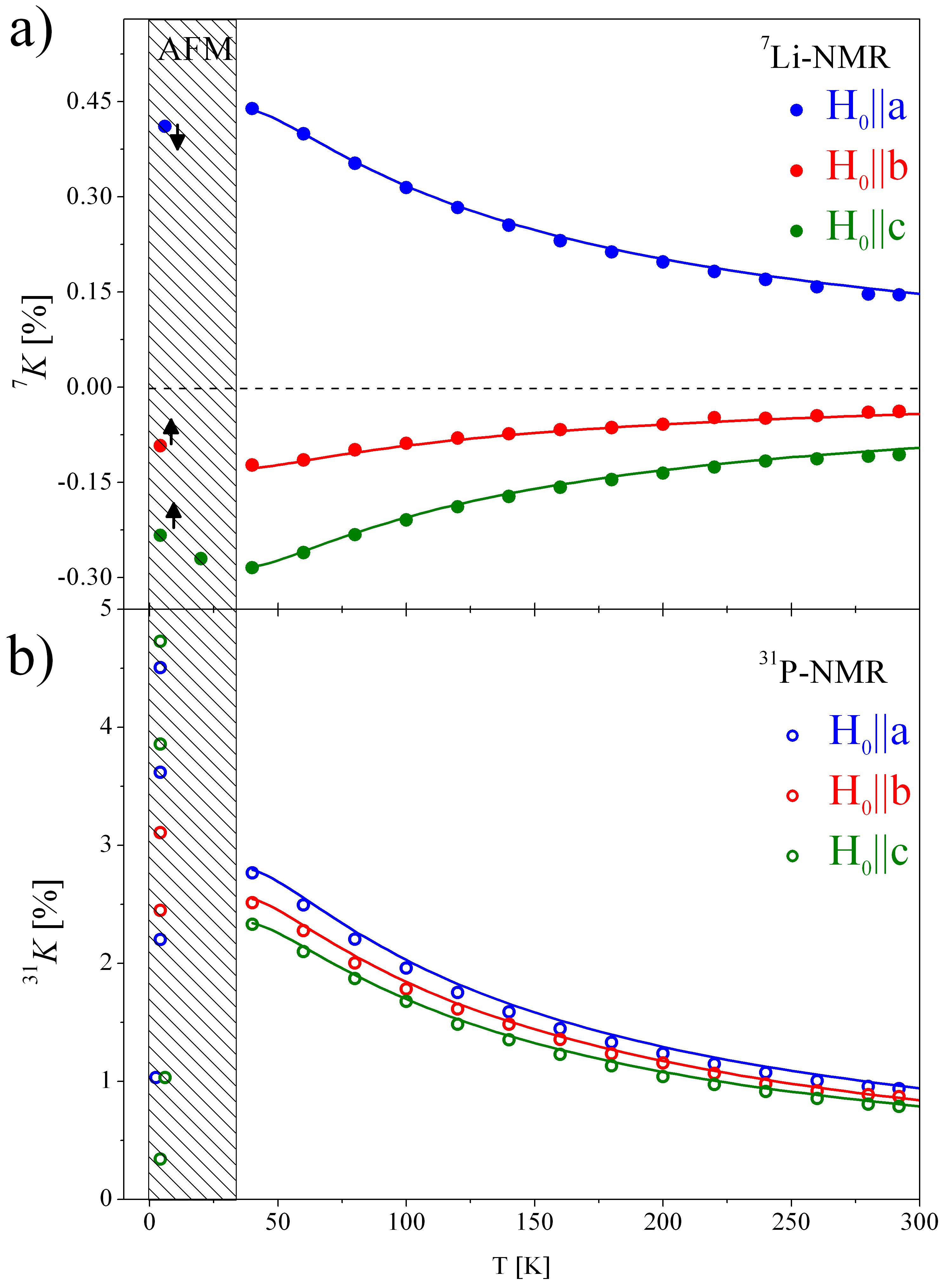}
  \caption{Temperature dependence of the Knight shift (a) $^7K$ and (b)
$^{31}K$ for the orientations $H_0\Arrowvert a$, $H_0\Arrowvert b$
and $H_0\Arrowvert c$ in the paramagnetic and AFM phase. The
measured data are indicated by the dots. The solid lines
correspond to the Knight shift which is calculated by
$K=A^\mathrm{exp}_{mm}\chi(T))$ with m=1,2,3 for $H_0\Arrowvert
a$, $H_0\Arrowvert b$ and $H_0\Arrowvert c$, respectively. The
diagonal tensor elements $A^\mathrm{exp}_{mm}$ are determined from
fits of Clogston Jaccarino plots and listed in
Tab.~\ref{tab:Tensors}. (a) The presence of the anisotropic
dipolar coupling leads to positive and negative $^7K$ for
corresponding orientation. In the AFM phase the single resonance
line shifts towards $^7K=0$ which is indicated by the arrows. (b)
The superposition of the field caused by the contact term and the
field of the dipolar coupling leads to a comparatively high
$^{31}K$. In the AFM phase four resonance lines are observed for
$H_0\Arrowvert a$ and $H_0\Arrowvert c$ and two for $H_0\Arrowvert
b$.}
  \label{fig:NMR_chi_abc_Li}
\end{figure}

The temperature dependence of the resonance frequencies for
$^7$Li- and $^{31}$P, $\nu (T)$, is shown in
Fig.~\ref{fig:NMR_chi_abc_Li} (a) and (b). In the paramagnetic
phase, $\nu (T)$ can be described by
\begin{equation}
  \nu(T)=\gamma_n H_0 (1+A_{mm} \chi(T)).
  \label{eqn:omega_T}
\end{equation}
where $A_{mm}$ are the diagonal dipolar hyperfine coupling tensor
elements for the corresponding orientations~($H_0\Arrowvert a$,
$H_0\Arrowvert b$, $H_0\Arrowvert c$, m=1, 2, 3 respectively). The
magnetic susceptibility is isotropic in the paramagnetic
phase~\cite{wizent2009}, and thus $\chi(T)$ is a scalar in
Eqn.~\ref{eqn:omega_T}. From Clogston Jaccarino plots the diagonal
elements $A^{exp}_{mm}$ are obtained for the corresponding
orientations (see Appendix Tab.~\ref{tab:Tensors}). Since $A_{22}$
and $A_{33}$ for $^7$Li are negative, the resonance lines for
$H\parallel a$ and $b$ shift to lower frequencies. For the
$^{31}$P, there is an additional temperature dependent contact
term (see Fig.~\ref{fig:NMR_abc_Li_P} (b)) which leads to a
positive shift of the resonance frequency for all three
directions. The contact term is also responsible for the large
deviation of the calculated and experimentally determined coupling
constants in Tab.~\ref{tab:Tensors}. Further deviations may arise
from the anomalous shape of the spectra due to the Mn disorder.

\begin{figure}[!ht]
  \centering
  \includegraphics[width=0.45\textwidth]{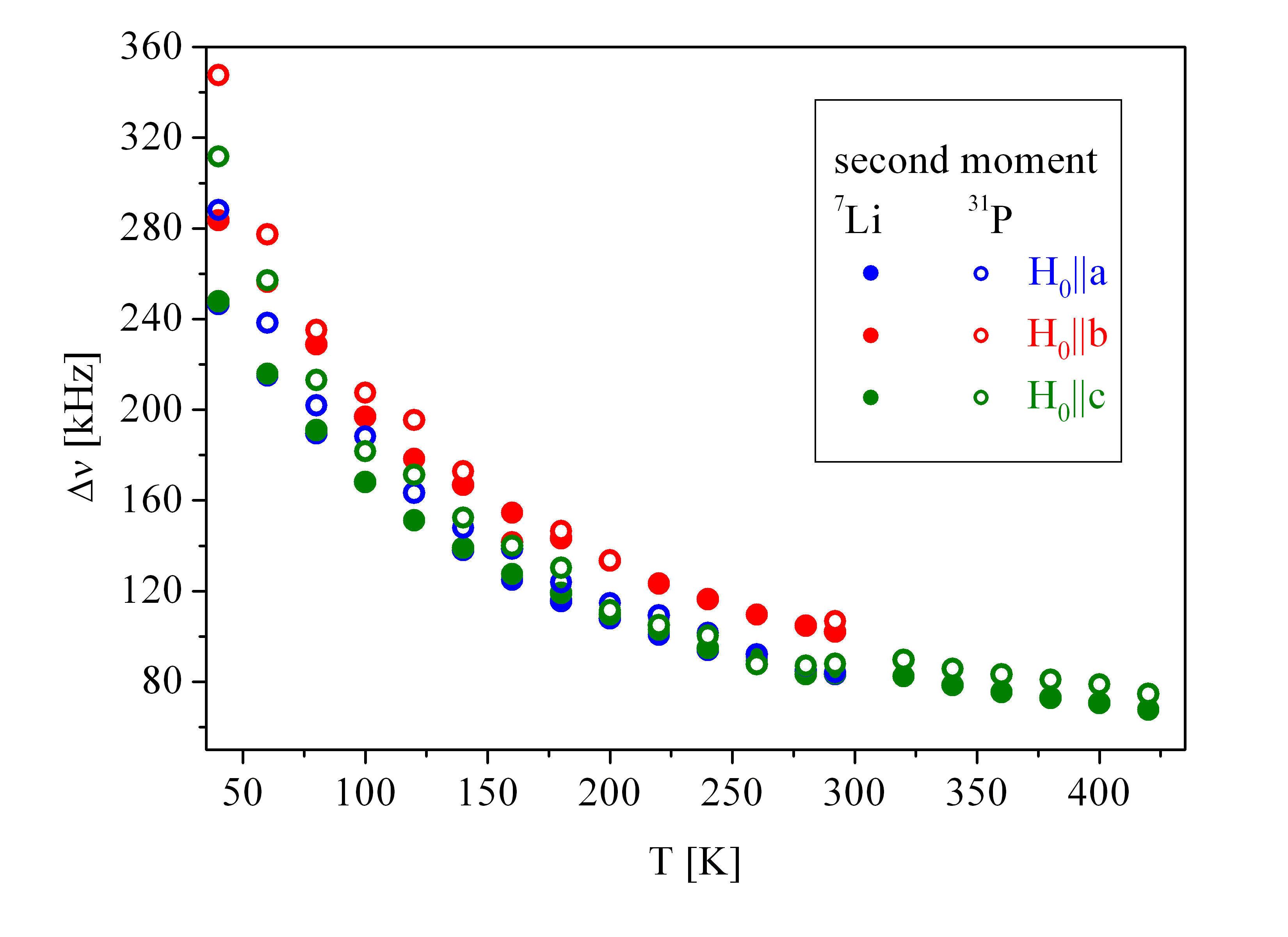}
  \caption{Comparison of the temperature dependent second moment $\Delta \nu$ between $^7$Li-
and $^{31}$P-NMR spectra.
The second moment of the \phos-spectra is almost always larger than the corresponding second moment of
the \li-spectra.
This effect is predominantly related to a distribution of internal fields by
the contact interaction on the P nuclei. For $H_0\Arrowvert
c$ the spectra are measured up to 420~K, where the second moment
remains almost constant at 80~kHz.}
\label{fig:FWHM_abc_Li_P}
\end{figure}

The temperature dependence of the square root of the second moment
$\Delta \nu$ of the $^7$Li and $^{31}$P-NMR spectra is shown in
Fig.~\ref{fig:FWHM_abc_Li_P}. With decreasing temperature $\Delta
\nu$ increases which is the expected behavior for magnetic
broadening which scales with the susceptibility $\chi$. For
$H_0\Arrowvert c$ the spectra were measured up to 420~K~(see
Fig.~\ref{fig:FWHM_abc_Li_P}). Above $\sim$300~K, the second
moment is almost constant yielding a value of $\Delta \nu \sim
80$~kHz. As can be seen in Fig.~\ref{fig:FWHM_abc_Li_P}, $\Delta
\nu$ of the \phos-NMR spectra is slightly larger for the
corresponding orientations than $\Delta \nu$ of the \li . The
reason for this could be a distribution of the hyperfine fields
transferred by the contact interaction which affects only the
\phos , but not the \li . The effect of the contact interaction is
isotropic, hence the \phos-NMR spectra are broadened without
additional features in the shape of the resonance line.

\subsection{Angle-dependent NMR frequencies} \label{sec:orient_dep}

The full experimental orientation dependence of $^7$Li- and
$^{31}$P-NMR can be described consistently with the susceptibility
data and the  hyperfine coupling tensor $\hat{A}^{para}_\mathrm{dip}$.

\begin{figure}[!t]
  \center
  \includegraphics[width=0.45\textwidth]{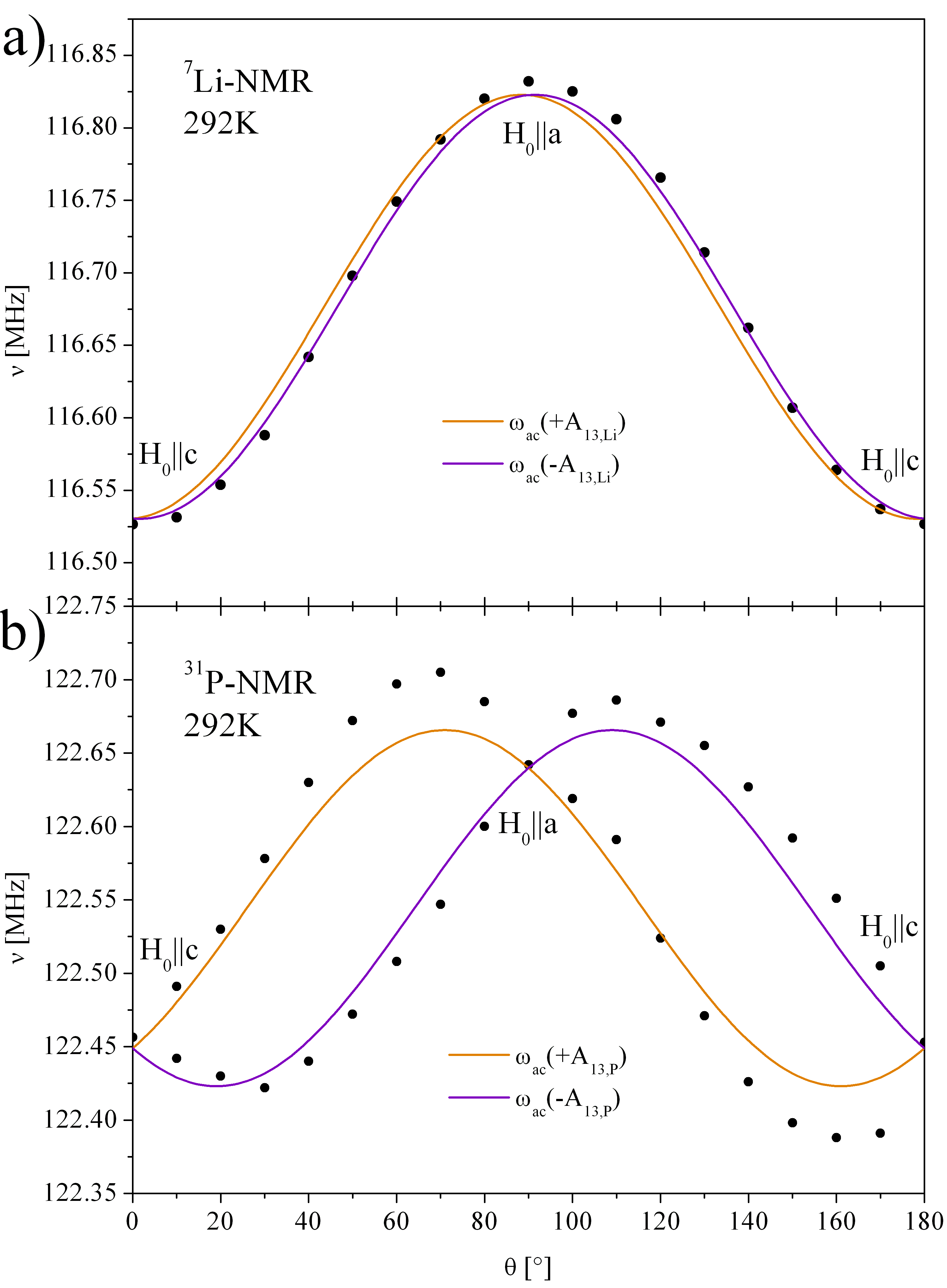}
  \caption{Experimental orientation dependence of $^7$Li- and $^{31}$P-NMR between $H_0\Arrowvert c$
and $H_0\Arrowvert a$.
Expected curves: $\nu_{ac}=\gamma_{n}H_0(1+(A_{11}\sin^2\theta +2A_{13}\sin\theta \cos\theta
+A_{33}\cos^2\theta)\chi(292K))$ for the corresponding $^7$Li and $^{31}$P tensor elements. The diagonal
elements are experimental, $A_{13}$ is calculated. a) The absolute value of
$\pm A_{13,Li}$ is relatively small which
leads to a small splitting of the $^7$Li-NMR resonance line on the order of $\sim$10~kHz which is not
oberserveable due to the broad resonance line of $\sim$100~kHz.
b) The line splitting of $\sim$200~kHz is observable. The calculated off diagonal element $\pm
A_{13,P}$ is $\sim$10~times larger than $\pm A_{13,Li}$ which increases the splitting in the
$^{31}$P-NMR spectrum.}
  \label{fig:Li_P_ac}
\end{figure}

\begin{figure}[!t]
  \center
  \includegraphics[width=0.45\textwidth]{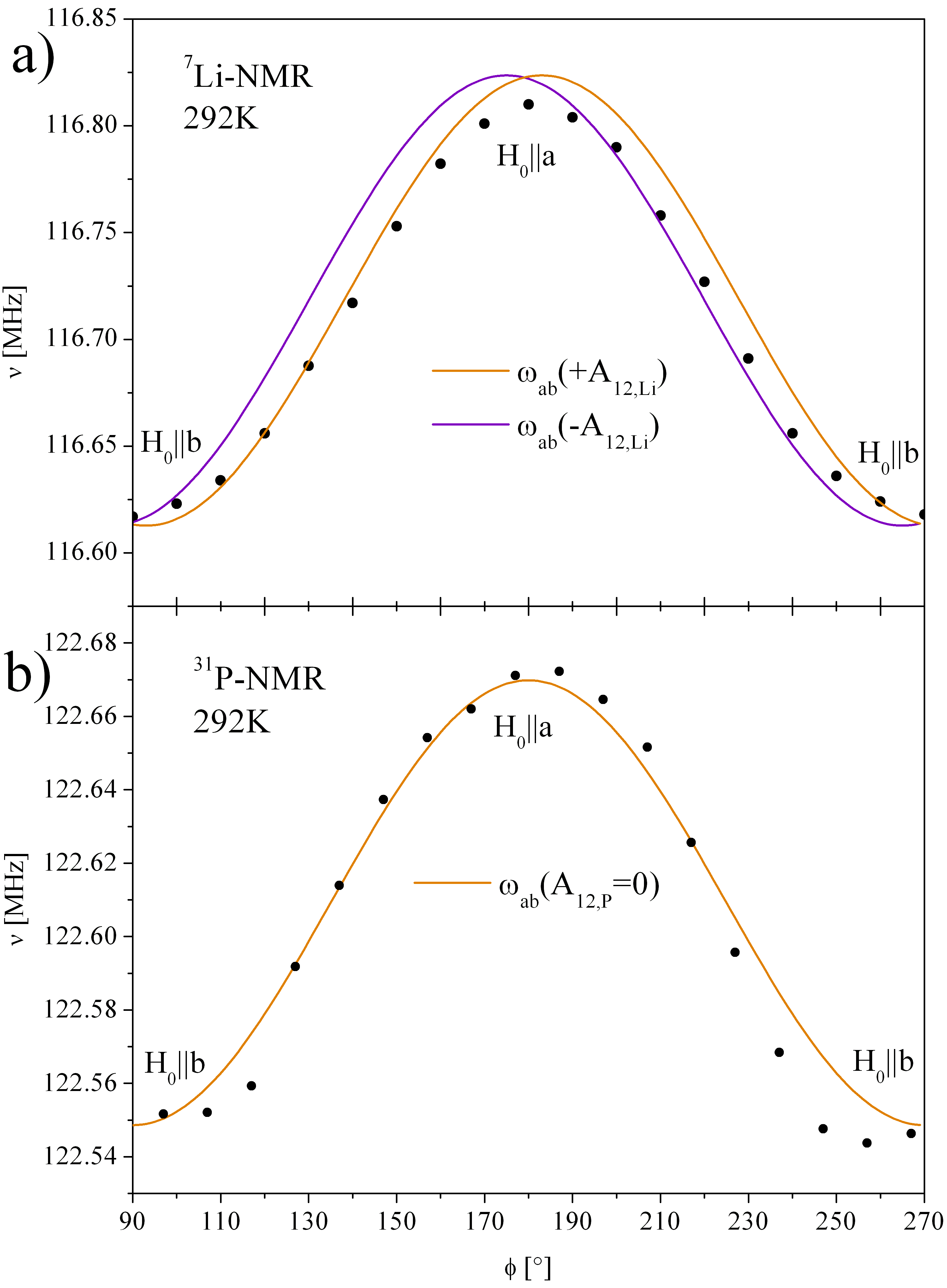}
  \caption{Experimental orientation dependence of $^7$Li- and $^{31}$P-NMR between $H_0\Arrowvert b$
and $H_0\Arrowvert a$. Expected curves:
$\nu_{ab}=\gamma_{n}H_0(1+(A_{11}\cos^2\phi +2A_{12}\sin\phi
\cos\phi +A_{22}\sin^2\phi)\chi(292K))$. a) $A_{12,Li}$ causes the
splitting on the order of $\sim$20~kHz which is too small to
observe it by $^7$Li-NMR because of the broad linewidth of
$\sim$100~kHz. b) $A_{12,P}$ vanishes~(see Appendix
Tab.~\ref{tab:Tensors}) which means that no splitting is present.}
  \label{fig:Li_P_ab}
\end{figure}

\begin{figure}[!t]
  \center
  \includegraphics[width=0.45\textwidth]{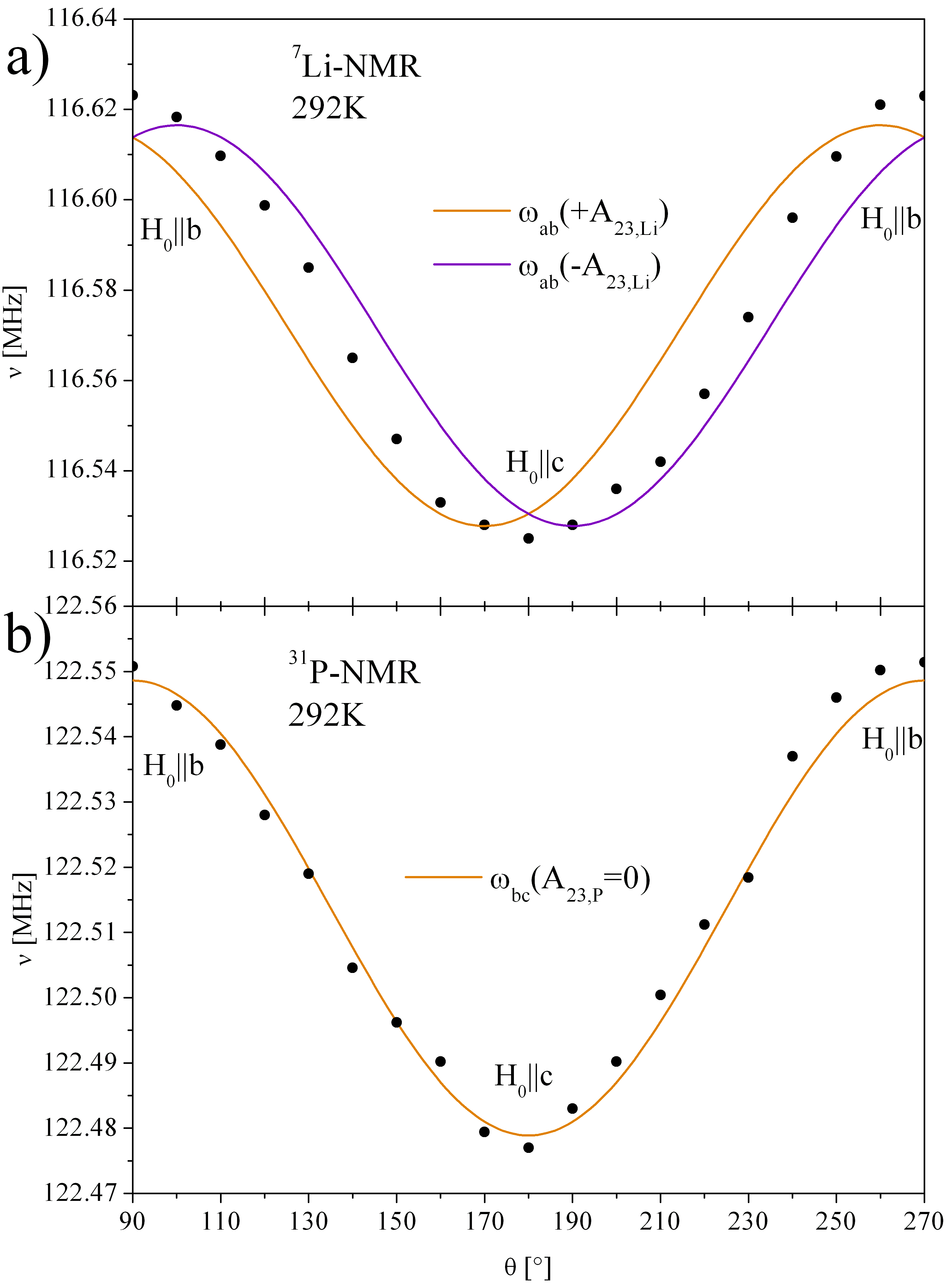}
  \caption{Experimental orientation dependence of $^7$Li- and $^{31}$P-NMR between $H_0\Arrowvert b$
and $H_0\Arrowvert c$. Expected curves:
$\nu_{bc}=\gamma_{n}H_0(1+(A_{22}\sin^2\theta +2A_{23}\sin\theta
\cos\theta +A_{33}\cos^2\theta)\chi(292K))$. a) $A_{23,Li}$ causes
the splitting on the order of $\sim$30~kHz which is too small to
observe it by $^7$Li-NMR because of the broad linewidth of
$\sim$100~kHz. b) $A_{23,P}$ vanishes~(see Appendix
Tab.~\ref{tab:Tensors}) which means that no splitting is present.}
  \label{fig:Li_P_bc}
\end{figure}

The angle-dependent NMR data was obtained by rotating the crystal
around the crystal axes from $H_0\Arrowvert c$ to $H_0\Arrowvert
a$, from $H_0\Arrowvert b$ to $H_0\Arrowvert a$, and from
$H_0\Arrowvert b$ to $H_0\Arrowvert c$ at 292~K. The angle
dependence at this temperature can be described by
\begin{equation}
\label{eqn:omega_orient}
\begin{split}
\nu(\phi,\theta)=&\gamma_nH_0(1+\\ &\begin{pmatrix}\cos\phi \sin\theta \\ \sin\phi \sin\theta \\
\cos\theta \end{pmatrix}\hat{A}^{para}_\mathrm{dip}\begin{pmatrix}\cos\phi \sin\theta \\ \sin\phi
\sin\theta \\ %\nonumber
\cos\theta \end{pmatrix}\chi(292K))
\end{split}
\end{equation}

where the orientation of the crystal structure in the external
field is expressed by the unit vectors given in spherical
coordinates. The experimental and calculated data according to
Eqn.~\ref{eqn:omega_orient} are shown in Fig.~\ref{fig:Li_P_ac} to
Fig.~\ref{fig:Li_P_bc}. For the calculated curves the
corresponding experimental diagonal elements and the calculated
off diagonal elements are used.
The splitting of the resonance lines for angles off the high
symmetry directions is caused by the different magnetic $^7$Li and
$^{31}$P sites~(see Appendix Tab.~\ref{tab:offelements}). The
different sites lead to positive and negative signs of the off
diagonal elements $A_{12}$, $A_{13}$, $A_{23}$.

In Fig.~\ref{fig:Li_P_ac} (a) the $^7$Li-NMR spectrum shows a
single resonance line in contrast to the $^{31}$P-NMR spectrum in
Fig.~\ref{fig:Li_P_ac} (b) which shows two resonance lines for
angles off $H_0\Arrowvert a$ and $H_0\Arrowvert c$. From
comparison of the angle dependent $^{31}$P-NMR data reported
in~\cite{Mays1963} and Fig.~\ref{fig:Li_P_ac} (b) the crystal axes
\textit{a} and \textit{c} are confirmed. For rotating between
$H_0\Arrowvert c$ and $H_0\Arrowvert a$,
Eqn.~\ref{eqn:omega_orient} simplifies to an equation which
includes only the tensor elements $A_{11}$, $A_{33}$, $A_{13}$.
The absolute value of $A_{13}$ determines the size of splitting
that means the difference between the violet and orange curve in
Fig.~\ref{fig:Li_P_ac}. In the $^7$Li-NMR spectra the $A_{13,Li}$
element leads to a splitting on the order of $\sim$10~kHz which is
completely covered by the broad linewidth of $\sim$100~kHz. The
value $A_{13,P}$ for $^{31}$P is $\sim$10~times larger than
$A_{13,Li}$~(see Tab.~\ref{tab:Tensors}) and leads to a splitting
of 200~kHz which can be nicely observed in the $^{31}$P-NMR
spectra. Fitting the experimental $^{31}$P-NMR orientation
dependent data in Fig.~\ref{fig:Li_P_ac} (b) leads to experimental
values for $A_{13,P}$ (see Appendix Tab.~\ref{tab:Tensors}).

The experimental results for rotating the crystal between
$H_0\Arrowvert b$ and $H_0\Arrowvert a$ are shown in
Fig.~\ref{fig:Li_P_ab}. For these orientations
Eqn.~\ref{eqn:omega_orient} simplifies to an equation including
only $A_{11}$, $A_{22}$ and $A_{12}$. For this case $A_{12}$
determines the size of the splitting.

The orientation dependence between $H_0\Arrowvert b$ and
$H_0\Arrowvert c$ is shown in Fig.~\ref{fig:Li_P_bc}. Here,
Eqn.~\ref{eqn:omega_orient} includes only the tensor elements
$A_{22}$, $A_{33}$ and $A_{23}$. In this case the off diagonal
element $A_{23}$ is responsible for the splitting.

\subsection{Discussion of the angle dependence} \label{sec:comparison}

The angle dependence of the NMR spectra provides another
indication that the broad lines and anomalous shape of the
resonance lines in Fig.~\ref{fig:NMR_abc_Li_P} are due to Mn
disorder, and not due to differently oriented domains of the
crystal. First of all, the angle dependencies of the \li\ and the
\phos\ are completely different, but the shapes of the spectra are
very similar. This can be seen in Fig.~\ref{fig:Li_P_ac}: whereas
the \phos\ lines split depending on the angle $\theta$, the \li\
lines do not. In contrast, the shape of the spectra for
$H_0\Arrowvert a$ in Fig.~\ref{fig:NMR_abc_Li_P} is very similar
for both isotopes. An angle dependence as that of the \phos\ in
Fig.~\ref{fig:Li_P_ac} (b) would clearly lead to a different shape
of the spectrum compared to the \li . Similar arguments hold for
the other two directions. Especially for $H_0\Arrowvert b$, the
shape of the spectra for \li\ and \phos\ are very similar, but the
angle dependencies in Fig.~\ref{fig:Li_P_ab} and
Fig.~\ref{fig:Li_P_bc} are not. For example, in
Fig.~\ref{fig:Li_P_ab}, the difference of the resonance frequency
of the \li\ for $H_0\Arrowvert a$ and $H_0\Arrowvert b$ is
$\sim$200~kHz, whereas it is only $\sim$120~kHz for the \phos . In
contrast, the distance between the shoulder and the peak for
$H_0\Arrowvert b$ in Fig.~\ref{fig:NMR_abc_Li_P} is very similar
for both isotopes, if not even larger for the \phos . The same
argument is true for the linewidths in general: the angle
dependent splitting is always larger for the \li , whereas the
FWHM tends to be larger for the \phos . A broadening from
differently oriented domains of a crystal would lead to the
opposite trend, as can be deduced from the angle dependent
measurements.

Furthermore, the comparison of the calculated and experimental
tensor elements indicates the presence of an additional hyperfine
interaction for the $^{31}$P nuclei, whereas the values for the
\li\ in principle agree: the diagonal element $A_\mathrm{11,Li}$
determined by NMR and the value determined from the crystal
structure is the same, and the values for $A_\mathrm{22,Li}$ and
$A_\mathrm{33,Li}$ are showing only a small residue. The
difference can be explained by the asymmetric shape of the
resonance line for $H_0\Arrowvert b$ which is most likely due to
the Mn disorder. The calculation of the resonance frequency from
the center of gravity therefore leads to a smaller shift of the
resonance frequency than expected from the calculated
$A_\mathrm{22,Li}$. The residue of $A_\mathrm{33,Li}$ possibly
arises from the Mn disorder, too. Therefore, it can be concluded
that the Mn disorder has its largest influence in \textit{b} and
\textit{c} direction because of the larger discrepancies of
$A_\mathrm{22,Li}$ and $A_\mathrm{33,Li}$ compared to
$A_\mathrm{11,Li}$. A small displacement of some of the Mn atoms
from their original positions in the lattice leads to a
distribution of the dipolar hyperfine couplings resulting in the
broad lines and the asymmetric shape. Note that the off diagonal
elements for $^7$Li can in principle be determined, too. However,
the splitting is only very small ($\sim$10-40~kHz), and therefore
can not be resolved within the broad resonance lines (width
$\sim$100~kHz).

In contrast to the \li , the comparison of the experimental and
calculated \textit{diagonal} hyperfine coupling elements for the
$^{31}$P shows large residues (see Appendix
Tab.~\ref{tab:Tensors}). Such a discrepancy has already been
reported in~\cite{Mays1963} and has been explained by non-dipolar
hyperfine interactions of the $^{31}$P nucleus with unpaired $3s$
(contact interaction) and/or $3p$ (core polarization) electrons of
the phosphorous. These unpaired $s$ and $p$ electrons arise from
the Mn-O-P-O-Mn superexchange path which is responsible for the
interlayer magnetic coupling \cite{LiPRB2009}. The difference
between the experimental and calculated resonance frequencies of
the $^{31}$P-NMR corresponds to an isotropic difference in the
diagonal elements of the hyperfine coupling tensor
of~0.4~T/$\mu_B$ which underlines the isotropic character which is
typical for a contact interaction.

The off diagonal element $A_{13,P}$ is causing a splitting of
$\sim$200~kHz for angles between $H_0\Arrowvert a$ and
$H_0\Arrowvert c$ which is larger than the averaged linewidth and
therefore observable by $^{31}$P-NMR. From the obtained data the
value of $A_{13,P}$ with positive and negative sign is extracted
by fitting both angle dependent $^{31}$P-NMR curves in
Fig.~\ref{fig:Li_P_ac} (b) by using Eqn.~\ref{eqn:omega_orient}.
The experimental absolute values of $\pm A_{13,P}$ are slightly
larger. The deviations may arise from the Mn disorder.

\subsection{NMR in the antiferromagnetic phase}

\begin{figure}[!t]
  \center
  \includegraphics[width=0.45\textwidth]{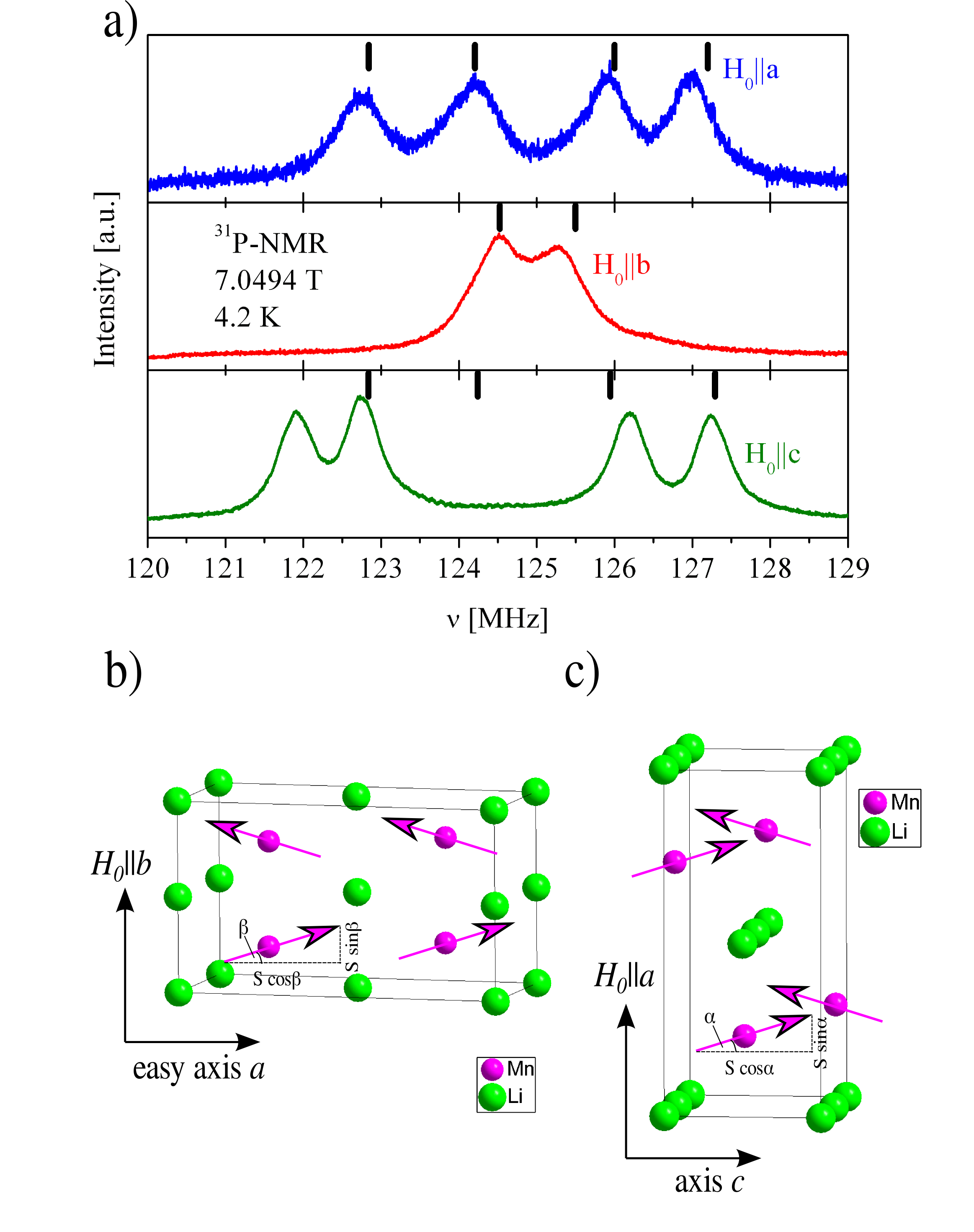}
  \caption{a) $^{31}$P-NMR spectra for $H_0\Arrowvert a$, $H_0\Arrowvert b$ and $H_0\Arrowvert c$ at
4.2~K. By asumming tilt angles, the $^{31}$P-NMR frequencies can be
calculated as shown in Tab.~\ref{tab:spin_afm}. The best match of experimental and calculated
frequencies is achieved for $\alpha=11^\circ$, $\beta=4^\circ$ and $\gamma=13^\circ$. The calculated
frequencies are indicated by the short black lines. b) Tilts of the electronic spins in the AFM state.
Exemplified shown for tilts in $b$ direction. For tilts in $c$ direction~(tilt angle $\gamma$) the
situation is similar.
c) Tilts in the SF state in the direction of $a$ by the angle $\alpha$. The tilt causes an additional
component of the spins in field
direction.}
  \label{fig:spin_arrows}
\end{figure}

In order to check the dipolar coupling and to investigate the
influence of the spin flop on the NMR spectra, we measured the
$^7$Li- and $^{31}$P spectra in the AFM phase at 4.2~K. In the
single crystal at hand, the spin flop field at this temperature
amounts to 3.9 T. The NMR spectra are shown in
Fig.~\ref{fig:spin_arrows} (a) for the orientations $H_0\Arrowvert
a$, $H_0\Arrowvert b$ and $H_0\Arrowvert c$. In the AFM phase the
bulk susceptibility can not be used to calculate the resonance
frequencies from the dipolar hyperfine coupling, because a part of
electron spins changes its orientation to the opposite direction.
In zero external magnetic field, the electron spins align
antiparallel along the crystallographic $a$-axis. For the NMR
experiments, a strong external field $H_0=7$ T has been applied
which leads to a tilt of the electron spins for the directions
$H_0\Arrowvert b$ and $H_0\Arrowvert c$ by the angle~$\beta$
and~$\gamma$, respectively, and to a spin-flop transition plus an
additional tilt by the angle~$\alpha$ for the direction
$H_0\Arrowvert a$ (see Fig.~\ref{fig:spin_arrows} (b) and (c))).
The occurrence of a spin-flop transition has been reported
in~\cite{Toft-PetersenPRB2012,ranicar1967} for $H_0>4$~T parallel
$a$. The tilt angle in the spin flop phase is labeled $\alpha$.

Eqn.~\ref{eqn:omega_AFM} considers the additional fields induced
by the contact interaction $B_{c}$ and the dipolar field
$B^\mathrm{AFM}_\mathrm{dip}$ which stems from the antiferromagnetic order of
the electron spins.
\begin{equation}
  \nu=\gamma_n(H_0+B_{c} + B^\mathrm{AFM}_\mathrm{dip})
  \label{eqn:omega_AFM}
\end{equation}

The dipolar field $B^\mathrm{AFM}_\mathrm{dip,Li}$ vanishes for
the $^7$Li sites in the AFM phase, therefore the $^7$Li-NMR
spectra show a single resonance line at the frequency of a bare
nucleus. In Fig.~\ref{fig:NMR_chi_abc_Li} (a) the expected
behavior is experimentally indicated by small shifts of the
$^7$Li-NMR resonance frequencies in the direction of the unshifted
frequency at 4.2~K. The incomplete shifting to $\gamma_{Li}H_0$
can be explained by a remaining dipolar field induced by the
tilted electron spins.

The resonance lines in the AFM phase of the $^{31}$P are shown in
Fig.~\ref{fig:spin_arrows} (a) for the orientations $H_0\Arrowvert
a$, $H_0\Arrowvert b$ and $H_0\Arrowvert c$. The number of
resonance lines in the $^{31}$P-NMR spectra equals the number of
different calculated dipolar fields
$B^\mathrm{AFM}_\mathrm{dip,P}$ (see Appendix
Tab.~\ref{tab:spin_afm}).

The additional component in the field direction due to the tilt of
the electronic spins is considered in Tab.~\ref{tab:spin_afm} for
the corresponding orientations. Then, the $^{31}$P NMR frequencies
are calculated as a function of the tilt angles. The best match of
experimental and calculated resonances are found for
$\alpha=11^\circ$, $\beta=4^\circ$, and $\gamma=13^\circ$. The
calculated frequencies are indicated by bars in
Fig.\ref{fig:spin_arrows}. For $H_0\Arrowvert a$ and
$H_0\Arrowvert b$ the experimental and calculated resonance
frequencies agree quite well, and the value for $H_0\Arrowvert a$
is also consistent with the tilt angle of 5.4$^\circ$ at
$H_0=4.5~$T determined by neutron
scattering~\cite{Toft-PetersenPRB2012} since the tilt angle of the
spins almost depend linearly on the field. Only for $H_0\Arrowvert
c$, the calculated resonances on the low frequency side for some
reason do not agree well with the experimental data.

In general, our results show that it is possible to determine the
magnetic structure by NMR in the AFM phase, and even to determine
the small tilt of the spins due to the applied magnetic field.

\section{Diffuse x-ray diffraction}

The NMR lineshape and angle dependence provide strong evidence for
the presence of disorder of the Mn-sublattice. In order to verify
this conclusion we performed diffuse HE-XRD
experiments on the same single crystals studied by NMR.
The  HE-XRD experiments were performed using a photon energy of 100 keV. At these energies the penetration depth of the x-rays is of the order of millimeters,
which guarantees the detection of true bulk properties and enables
a direct comparison to the NMR results.
\begin{figure}[t!]
  \centering
  \includegraphics[width=\columnwidth]{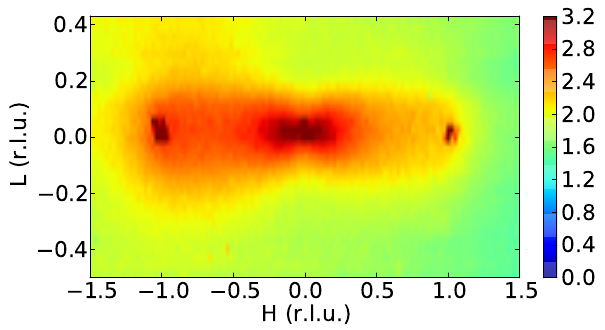}
  \caption{Diffuse scattering intensity around the forbidden (030) position. The intensity is given in
counts/second at 100~mA ring current.} \label{fig:HL_030}
\end{figure}

A representative data set is displayed in Fig.~\ref{fig:HL_030},
where the diffuse scattering around the symmetry forbidden (030)
is displayed. The very broad and diffuse HE-XRD intensity in the
HL-plane can clearly be observed. The strong diffuse scattering
reveals significant structural disorder in the studied single
crystals. The HE-XRD intensity is proportional to
$(\mathbf{Q}\cdot \mathbf{d})^2$, where $\mathbf{Q}$ is the
scattering vector and $\mathbf{d}$ represents the shift of a
lattice site away from its ideal position. The diffuse scattering
around $\mathbf{Q}$=(0,3,0) therefore implies that the lattice
disorder involves shifts parallel to the $b$-axis, which are only
very weakly correlated along $a$. The distribution of the diffuse
intensity is also very anisotropic within the (H3L)-plane, i.e.,
the underlying disorder is anisotropic in real space, consitent
with the NMR results.
The data in Fig.~\ref{fig:HL_030} further implies that the
symmetry of the ideal lattice is not strictly fulfilled and
therefore the symmetry forbidden ($\pm$1,3,0) and (0,3,0)
reflections are also observed. In addition to this the HE-XRD
intensity is dominated by the shifts of the heavy elements in the
lattice. The results in Fig.~\ref{fig:HL_030} therefore directly
reveal the disorder in the Mn-sublattice and is in perfect
agreement with the conclusions drawn from the NMR data.

Based on the available HE-XRD data, we cannot characterize the
structural disorder in more detail. This is subject of ongoing
investigations.

\section{Conclusions}

We presented angle and temperature dependent $^7$Li- and
$^{31}$P-NMR and diffuse x-ray diffraction measurements in a
LiMnPO$_4$ single crystal.
The $^7$Li- and $^{31}$P-NMR resonance lines in the AFM
and SF phase are consistent with the calculated resonances, when
tilts of the magnetic moments in the direction of the external
field are assumed.
The broad $^7$Li- and $^{31}$P-NMR resonance lines in the
paramagnetic phase indicate the existence of Mn site disorder. In
combination with the susceptibility data the diagonal elements of
the hyperfine coupling tensor are determined and compared with the
values calculated from the crystal structure.
The highest deviations show $A_\mathrm{22,Li}$ and $A_\mathrm{33,Li}$ which
leads to the conclusion that the Mn disorder exists predominantly
in $b$ and $c$ direction. This is fully consistent with our HE-XRD
data which indicates substantial-Mn disorder along the $b$
direction.
Interestingly, recent theoretical results imply a strong coupling
of the Li- and the Mn-sublattice in the fully lithiated
limit\,\cite{AsariPRB2011}. The Mn-disorder may hence be caused by
the distribution of the mobile Li-ions within the MnPO$_4$ host.
Our finding of Mn disorder in \lmpo\ therefore provides
experimental evidence that the movement of Li within LiMnPO$_4$ is
significantly coupled to lattice distortions. Certainly, this will
affect also the mobility of the Li in \lmpo , and therefore the
performance of this material as a battery.

\section*{Acknowledgments}
This research was financially supported by the DFG priority
program SPP1473 (Grant No. GR3330/3-1 and KL1824/5). N.W. and R.K.
acknowledge support by the BMBF via project 03SF0397. J.G. and
S.P. gratefully acknowledge the financial support by the German
Research Foundation through the Emmy-Noether program (Grant No.
GE1647/2-1).

\section*{Appendix: Calculated and Experimental Hyperfine Couplings}

\begin{table*}
    \center
    \caption{Comparison of calculated and NMR determined tensor elements for the paramagnetic phase. The
elements for the AFM phase are calculated. All elements are given
in T/$\mu_B$.}
    \begin{tabular}{cc|ccc|c}
     Nuclei & Tensor Element & \multicolumn{3}{c|}{Paramagnetic phase} & AFM phase\\
      &  & Crystal structure & NMR & Residue & Crystal structure\\
     & & (calculated) & (experimental) & & (calculated)\\ \hline
     & & & & & \\
     $^7$Li & $A_\mathrm{11,Li}$ & 0.0690 & 0.0717 & 0.0027 & 0\\
     &$A_\mathrm{22,Li}$ & -0.0401 & -0.0198 & 0.0203 & 0\\
     &$A_\mathrm{33,Li}$ & -0.0289 & -0.0430 & -0.0141 & 0\\
     &$A_{12,Li}$ & $\pm0.0058$ & - & - & 0\\
     &$A_{13,Li}$ & $\pm0.0030$ & - & - & 0\\
     &$A_{23,Li}$ & $\pm0.0060$ & - & - & 0\\
     & & & & & \\ \hline
     & & & & & \\
     $^{31}$P & $A_\mathrm{11,P}$ & 0.0376 & 0.4429 & 0.4053 & $\pm0.0881$\\
     &$A_\mathrm{22,P}$ & -0.0193 & 0.3847 & 0.4040 & $\pm0.1560$\\
     &$A_\mathrm{33,P}$ & -0.0183 & 0.3710 & 0.3893 & $\pm0.0679$\\
     &$A_{12,P}$ & 0 & - & -& 0\\
     &$A_{13,P}$ & $\pm0.0282$ & +0.0388~/~-0.0362 & 0.0106~/~0.0080 & $\pm0.0374$\\
     &$A_{23,P}$ & 0 & - & - & 0
   \end{tabular}
   \label{tab:Tensors}
\end{table*}

\begin{table*}[!hbt]
    \center
    \caption{The signs for the off diagonal elements of the magnetic unequal $^7$Li and $^{31}$P sites
are shown. The sites are given in fractional coordinates.}
    \begin{tabular}{l|ccc||cc|ccc}
     $^7$Li sites & $A_{12,Li}$ & $A_{13,Li}$ & $A_{23,Li}$ & & $^{31}$P sites & $A_{13,P}$\\
     & & & & & &\\ \hline
     & & & & & &\\
     (0 / 0 / 0) & + & +  & - &  & (0.907 / 0.75 / 0.591) & -\\
     (0 / 0.5 / 0) & - & +  & + & & (0.092 / 0.25 / 0.408) & -\\
     (0.5 / 0 / 0.5) & + & - & + & & (0.407 / 0.75 /0.906) & +\\
     (0.5 / 0.5 / 0.5) & - & - & - & & (0.592 / 0.25 / 0.094)& +\\
   \end{tabular}
   \label{tab:offelements}
\end{table*}

\begin{table*}[!b]
    \center
    \caption{The mathematical description for the orientations of $\vec{I}$ and $\vec{S_j^T}$ in the AFM
phase is given in the second column. The result is shown in the
third column. The number of different
$B^\mathrm{AFM}_\mathrm{dip,P}$ is obtained from the combination
possibilities of the $\pm$~sign of the tensor elements in the
third column. The number of different dipolar fields
$B^\mathrm{AFM}_\mathrm{dip,P}$ equals the number of experimental
$^{31}$P-NMR resonance lines. The tilt angles $\alpha$, $\beta$,
$\gamma$ are defined in Fig.~\ref{fig:spin_arrows} (b) and (c).}
    \begin{tabular}{c|ccc|c|c}
      & $\vec{I} \cdot$ & $\hat{A}_\mathrm{dip,P}^\mathrm{AFM}$ & $\cdot \vec{S}^T$ &
$B^\mathrm{AFM}_\mathrm{dip,P} \propto \sum_j
(\vec{I} \cdot \hat{A}^\mathrm{AFM}_{j,P} \cdot \vec{S}_j^T)$ & Number of \\
      & & & & & different $B^\mathrm{AFM}_\mathrm{dip,P}$\\ \hline
      & & & & &\\
      $H_0\Arrowvert a$ & $\begin{pmatrix} \frac{1}{2} \\ 0 \\ 0 \end{pmatrix}$ & $\begin{pmatrix} \pm
A_{11} & 0 & \pm A_{13} \\ 0 & \pm A_{22} & 0\\ \pm A_{13} & 0 &
\pm A_{33} \end{pmatrix}$& $\frac{5}{2}
\begin{pmatrix} \sin\alpha \\ 0 \\ \cos\alpha \end{pmatrix}$ & $\frac{5}{4} [(\pm A_{11})\sin\alpha +
(\pm A_{13})\cos\alpha]$& 4\\
      & & & &\\
      $H_0\Arrowvert b$ & $\begin{pmatrix} 0 \\ \frac{1}{2} \\ 0 \end{pmatrix}$ & $\begin{pmatrix} \pm
A_{11} & 0 & \pm A_{13} \\ 0 & \pm A_{22} & 0\\ \pm A_{13} & 0 &
\pm A_{33} \end{pmatrix}$& $\frac{5}{2}
\begin{pmatrix} \cos\beta \\ \sin\beta \\ 0 \end{pmatrix}$ & $\frac{5}{4} (\pm A_{22}) \sin\beta$&
2\\
      & & & &\\
     $H_0\Arrowvert c$ & $\begin{pmatrix} 0 \\ 0 \\ \frac{1}{2} \end{pmatrix}$ & $\begin{pmatrix} \pm
A_{11} & 0 & \pm A_{13} \\ 0 & \pm A_{22} & 0\\ \pm A_{13} & 0 &
\pm A_{33} \end{pmatrix}$ & $\frac{5}{2}
\begin{pmatrix} \cos\gamma \\ 0 \\ \sin\gamma \end{pmatrix}$ & $\frac{5}{4}[(\pm A_{13})\cos\gamma +
(\pm
A_{33})\sin\gamma]$& 4\\
    \end{tabular}
   \label{tab:spin_afm}
\end{table*}
\clearpage
\bibliographystyle{apsrev4-1}
\bibliography{bib_Rudisch_LiMnPO4}

\end{document}